\documentclass[manuscript]{acmart}

\usepackage{graphicx}

\AtBeginDocument{%
  \providecommand\BibTeX{{%
    \normalfont B\kern-0.5em{\scshape i\kern-0.25em b}\kern-0.8em\TeX}}}

\setcopyright{rightsretained}
\copyrightyear{2023}
\acmYear{2023}
\acmDOI{}

\acmConference[CHI 2023]{Hamburg '23: ACM Conference on Human Factors in Computing Systems}{April 23 -- April 28, 2023}{Hamburg, Germany}
\acmBooktitle{CHI 2023: Workshop on Trust and Reliance in AI-Human Teams,
  April 23 -- April 28, 2023, Hamburg, Germany}

\begin{document}

\title[Distrust in (X)AI]{Distrust in (X)AI - Measurement Artifact or Distinct Construct?}

\author{Nicolas Scharowski}
\authornote{Corresponding Author.}
\email{nicolas.scharowski@unibas.ch}
\orcid{0000-0001-5983-346X}

\author{Sebastian A. C. Perrig}
\email{sebastian.perrig@unibas.ch}
\orcid{0000-0002-4301-8206}

\affiliation{
  \institution{\\Center for General Psychology and Methodology, University of Basel}
  \streetaddress{Missionsstrasse 62a}
  \city{Basel}
  \postcode{CH-4055}
  \country{Switzerland}
}


\begin{abstract}

Trust is a key motivation in developing explainable artificial intelligence (XAI). 
However, researchers attempting to measure trust in AI face numerous challenges, such as different trust conceptualizations, simplified experimental tasks that may not induce uncertainty as a prerequisite for trust, and the lack of validated trust questionnaires in the context of AI. While acknowledging these issues, we have identified a further challenge that currently seems underappreciated - the potential distinction between \emph{trust} as one construct and \emph{distrust} as a second construct independent of trust. While there has been long-standing academic discourse for this distinction and arguments for both the one-dimensional and two-dimensional conceptualization of trust, distrust seems relatively understudied in XAI. In this position paper, we not only highlight the theoretical arguments for distrust as a distinct construct from trust but also contextualize psychometric evidence that likewise favors a distinction between trust and distrust. It remains to be investigated whether the available psychometric evidence is sufficient for the existence of distrust or whether distrust is merely a measurement artifact. Nevertheless, the XAI community should remain receptive to considering trust \emph{and} distrust for a more comprehensive understanding of these two relevant constructs in XAI.

\end{abstract}



\maketitle

\section{Introduction}

Trust has been studied for decades under different disciplinary lenses, such as philosophy \cite{fukuyama1996trust}, social sciences \cite{gambetta2000can}, and economics \cite{berg1995trust}. This has led to a multi-layered perspective on trust and sometimes divergent conceptions of trust in different disciplines. In the social sciences, trust has been defined as the expectation of non-hostile behavior; in the context of economics, trust is often conceptualized through game theory; in psychological terms, trust represents cognitive learning from experiences, and philosophically speaking, trust is based on moral relationships between individuals \cite{andras2018trusting}. Researchers have introduced accounts of trust that are appropriate in interactions between humans and machines \cite{lee2004trust}, including AI systems \cite{jacovi2021formalizing}. More recently, research on explainable AI (XAI) has regarded trust as a key motivation when creating more transparent and interpretable AI \cite{lipton2018mythos}.

While there seems to be a consensus in the XAI community that trust is a critical factor in human-AI interaction, researchers have identified challenges in measuring trust in the context of AI. For example, different conceptualizations of trust exist that are not clearly distinguished from one another (e.g., appropriate trust \cite{hoffman2018explaining}, calibrated trust \cite{langer2021we}, warranted trust \cite{jacovi2021formalizing}, or reliance \cite{poursabzi2018manipulating}). These various conceptualizations may lead to differences in the operationalization of trust. For example, \emph{trust as an attitude} should be viewed as a psychological construct and therefore be measured with questionnaires, whereas \emph{reliance as a behavior} can be measured with observational methods such as such as behavioral changes \cite{scharowski2022trust}. Moreover, researchers mostly measure trust at a single point in time that does not meaningfully capture the dynamics of trust development over time \cite{lewicki2006models}, and empirical studies often use proxy-tasks rather than actual decision-making tasks \cite{buccinca2020proxy} that do not necessarily impose uncertainty or risk that has been proposed as a prerequisite of trust \cite{jacovi2021formalizing}. Even if these challenges are addressed in the research design, AI-researchers who attempt to measure trust have to resort to questionnaires or survey scales from other disciplines because a validated questionnaire for trust in AI does not exist. One frequently used scale in XAI research is the \emph{Trust between People and Automation} (TPA) scale developed by \citet{jian2000foundations}. However, as the TPA was initially designed to examine trust in automation rather than trust in AI, researchers must rephrase the TPA items, such as "the system is dependable" to "the artificial intelligence is dependable." Terminological differences (e.g., using the word "AI" rather than "algorithm") impact people's perceptions and evaluations of technology \citep{langer_CHI_22}, which raises concerns about whether an adapted scale still measures what it was initially intended to measure. As a consequence, the psychometric quality of a scale (i.e., reliability and validity) must be reevaluated after such adaptations \citep{furr2011scale,juniper2009modified}. 

While it has been recognized that these issues complicate the measurement of trust in AI, we identified a different, noteworthy, and often underappreciated challenge - the lack of a theoretical consideration and potential distinction between \emph{trust} on the one hand and \emph{distrust} on the other. A first indication of this distinction was provided by \citet{spain2008towards}, who validated the TPA in the context of automation and crucially revealed a two-factor model in a factor analysis that distinguished trust and distrust. These results suggest that trust is not uni-dimensional as initially proposed by \citet{jian2000foundations}, but a two-dimensional construct consisting of trust and distrust as distinct factors \cite{spain2008towards}. Preliminary results from our ongoing work support this conceptualization and also indicate that a two-factor solution is preferable for the TPA in the context of AI \cite{perrig2023trust}. 
These findings have implications for using the TPA. Researchers have to consider the revealed two-factor structure and potential distinction between trust and distrust to obtain a better model fit and improved values for the scale in terms of validity and reliability compared to a single-factor conceptualization of trust. However, this raises the question of whether evidence of a two-factor structure is sufficient for two distinct and independent dimensions of trust and distrust or merely a research artifact caused by reverse-coded items.

The XAI community has provided important insights into how trust in AI can be developed and maintained, but distrust has been relatively understudied.\footnote{Indicated in a search on the ACM Digital Library on the 15. February 2023, searching for the words trust/distrust in combination with AI or XAI in the keywords and abstracts. For instance, at CHI 22, there were 12 publications on trust and AI/XAI, while there were no publications on distrust and AI/XAI. Overall, there were 280 hits in the ACM Digital Library for trust, in contrast to just six for distrust.} This ignores decades of research, which has been in a critical discourse on whether trust and distrust constitute the same construct at opposite ends of a continuum or should be treated as separate constructs on two distinct dimensions. A notable exception is the work from \citet{schelble2022addressing}, which appeared at the TRAIT 2022 workshop and partially motivated this position paper. 

Our contribution to the TRAIT Workshop 2023 is threefold: First, we provide a concise but informative overview of the theoretical discourse and the main arguments for distrust as a distinct construct. Second, we critically discuss from a psychometric perspective if the existence of a two-factor structure of the TPA is sufficient evidence for distrust or whether a two-factor solution might be a measurement artifact caused by reverse-coded items. Third, we critically discuss the possible implications of distrust as a distinct construct for XAI and suggest opportunities for future research examining whether distrust genuinely exists independently from trust. If this is the case, this contributes to a more comprehensive understanding of trust \emph{and} distrust in the context of AI that could inform the XAI community.

\section{Trust and distrust: Polar opposites, or independent constructs?}

According to \citet{lewicki2006models} there generally are two conceptualizations of trust.

\begin{itemize}
    \item[\textbf{I}] The uni-dimensional model, which treats trust and distrust as bipolar opposites, ranging from distrust to trust \citep[e.g.,][]{jian2000foundations, schoorman2007integrative, rotter1980interpersonal}.
    \item[\textbf{II}] The two-dimensional model, which argues that trust and distrust are two distinctly differentiable dimensions that can vary independently, each ranging from low to high \citep[e.g.,][] 
    {lewicki2006models, luhmann1979trust, sitkin1993explaining, saunders2014trust, McKnight2001trust, ou2009trust}.
\end{itemize}

Over the last 40 years, there have been advocates for both the uni-dimensional and two-dimensional models. In the following, we will introduce and mainly focus on the arguments for the two-dimensional trust model as we view this as the less established model. 

The underlying question raised by these two models is whether it is conceivable that trust and distrust can exist simultaneously and independently or whether trust and distrust are two sides of the same coin \cite{chang2013antecedents}. The uni-dimensional model suggests that high trust equals low distrust, and low trust equals high distrust, implying that the manifestation of trust is always dependent on the manifestation of distrust \cite{chang2013antecedents}. However, from the perspective of the two-dimensional model, distrust is more than the absence of trust, and vice versa \cite{kroeger2019unlocking}. Thus, high trust does not automatically imply low distrust, and the two constructs can coexist simultaneously and independently.

\citet{luhmann1979trust} is considered one of the main contributors to the theoretical foundation for a distinction between trust and distrust, and many later proponents of the two-dimensional model draw on his reasoning \cite{McKnight2001trust, kroeger2019unlocking, lewicki1998trust}. \citeauthor{luhmann1979trust} argued that distrust is associated with stronger emotional reactions than trust and reflects the negatively charged human survival instinct, while trust is more calm and composed, rendering the two constructs distinct. \citet{lewicki1998trust} further developed this idea and claimed that trust is based on more positive emotional responses (e.g., hope, faith, confidence) and distrust on more negative emotions (e.g., fear, skepticism, cynicism), so they may not just be at different ends of the same continuum, but orthogonal \cite{McKnight2001trust}. Based on these emotional differences, \citeauthor{lewicki1998trust} proposed a 2x2 framework with trust on one y-axis and distrust on the x-axis to capture not only the positive and negative emotions but also the potential for high and low levels of each construct to coexist simultaneously. Within this 2x2 framework, each quadrant represents one possible combination of the two constructs: low trust/low distrust (quadrant 1), high trust/low distrust (quadrant 2), low trust/high distrust (quadrant 3), and high trust/high distrust (quadrant 4), with each quadrant characterizing a distinct relationship between the two constructs and different challenges that go with it \cite{lewicki1998trust}. More recently, neuroimaging studies have supported this proposed difference in the emotional makeup of trust and distrust by showing that trust is more associated with the brain's reward, prediction, and uncertainty area. In contrast, distrust is associated with the brain's intense emotions and fear of loss areas, suggesting different neurological processes \cite{dimoka2010does}. 

The two-dimensional model (i.e., the potential coexistence of trust and distrust) is also supported by findings that attitudes can be ambivalent and possess both positive and negative components \citep{priester1996gradual}. For example, smokers trying to quit smoking may have both positive and negative feelings towards cigarettes, suggesting that positive and negative attitudes can coexist simultaneously \citep{cacioppo1994relationship}. This coexistence of apparently contradictory emotions allows for a more complex view of the trust relationship \cite{lewicki2006models}, acknowledging that there may be reasons to both trust and distrust simultaneously within the same relationship \cite{lewicki1998trust}. For example, A trusts B to do Y, yet distrusts B to do Z. \citet{McKnight2001trust} exemplified this with the cooperation between Stalin and Roosevelt in the second world war, where both parties trusted and distrusted each other at the same time. For \citeauthor{luhmann1979trust}, trust and distrust are "functional equivalents," meaning that rational actors use both trust and distrust to contain and manage uncertainty and complexity, but they do so by different means \cite{lewicki1998trust}. Trust reduces complexity by compelling a person to take action that exposes them to risk (i.e., undesirable outcomes are removed from consideration to form positive expectations \cite{kroeger2019unlocking}), while distrust reduces complexity by compelling a person to take protective action to reduce risk (i.e., undesirable outcomes are accentuated in consideration to form negative expectations \cite{kroeger2019unlocking}) \cite{benamati2006trust}. In summary, an argument can be made that both the antecedents (e.g., the associated emotions) and the consequences (e.g., the resulting function) of trust and distrust may be distinct \cite{cacioppo1994relationship, lewicki1998trust, mcknight2001while, chang2013antecedents}

This extensive work on the two-dimensional model, which for the purpose of this position paper can only be covered briefly, has led some authors to note that "most trust theorists now agree that trust and distrust are separate constructs that are the opposites of each other" \citep[p. 42]{McKnight2001trust}. However, there are also influential contributions and compelling arguments for the one-dimensional model. For example, \citet{schoorman2007integrative} replied to the statement above that "if they [trust and distrust] are opposites of each other, there is little added value to treating them as separate constructs" \citep[p. 8]{schoorman2007integrative}. The authors further noted that there is no theoretically or empirically viable evidence that trust and distrust are conceptually different and that researchers who studied distrust have merely reverse-coded measures of trust to represent their measures of distrust \cite{schoorman2007integrative}. This remark points to a psychometric question, namely, whether the two-factor solution found by \citet{spain2008towards} for the TPA questionnaire by \citet{jian2000foundations}, which our preliminary findings confirmed in the context of AI \cite{perrig2023trust}, is sufficient evidence for the independent existence of distrust or whether it is merely a research artifact.

\section{Distrust: Merely a research artifact?}

In psychometrics (i.e., a branch of psychology concerned with the theory and technique of measurement), there is evidence that uni-dimensional models are, at times, mistakenly considered multidimensional due to errors or artifacts of measurement \citep{schmitt1985factors}. In the following, we will illustrate how these errors and artifacts can be introduced into questionnaires.

A questionnaire or survey scale typically consists of a list of questions, called items, that reflect on different aspects of the underlying construct(s). When measuring a construct indirectly through the items of a scale, researchers make a crucial assumption: The response to the items is caused by the strength or level of the underlying construct \cite{devellis2017scale}. Thus, the construct that is not directly observable and its magnitude influences people's responses to the scale items.
During scale development and refinement, researchers can use exploratory and confirmatory factor analyses to identify and confirm theoretical models that best fit the data that was gathered using the scale's items \cite{brown2015confirmatory}. As part of this process, researchers form a theoretical model for their scale by defining and refining how many constructs are measured through the items and how these constructs relate to each other. Results from these processes thus shape how researchers understand their constructs of interest (e.g., trust and distrust) and how to use their scale for measurement in research.

However, past research has shown that so-called \emph{reverse-coded items} can distort the factor structures of scales \cite{pilotte1990impact,schmitt1985factors,schriesheim1981controlling,zhang2016examining}, thus leading to false conclusions regarding the dimensionality and theoretical structure of a questionnaire.
Reverse-coded items, also called negatively worded items, are items worded opposite to the regular scale items (i.e., positively formulated), which need to be re-coded prior to data analysis (e.g., a value of two on a Likert-type scale from 1 - 7 will be coded into a value of six).
Two common strategies used to create reverse-coded items are negating the target expression (e.g., adding "not") or working with antonyms (e.g., "bad" instead of "good") \citep{suarez2018using}.
For example, an XAI researcher developing a scale to measure trust could thus decide to create a reversed version of the item "The AI-system is reliable" \cite{jian2000foundations} through negation ("The AI-system is not reliable") or an antonym ("The AI-system is unreliable").

Theoretically, respondents who agree with a regular item should similarly disagree with a reverse-coded item. As a result, the negatively worded items should yield comparable results to the regular items after re-coding. Nevertheless, past research has cast doubt on this theoretical assumption.
Respondents are likely to misinterpret or misrespond to reverse-coded items \cite{sauro2011positive}, either by responding carelessly or because of so-called \emph{reversal ambiguity} \cite{weijters2012misresponse}.
Participants sometimes develop patterns of answering questionnaires based on the first few items they read (e.g., continuously responding with the value four), which reduces their attention \cite{schmitt1985factors}. The resulting carelessness leads respondents to overlook the negative wording of the reverse-coded items when filling out the questionnaire. \citet{schmitt1985factors} found that if just 10\% of respondents are careless while filling out a questionnaire, the factor structure can be distorted, highlighting the gravity of careless responding.

In addition, the wording used to create reverse-coded items can be ambiguous to the respondents if it leaves room for interpretation (i.e., display "reversal ambiguity"), causing even those who are careful while filling out the questionnaire to misrespond \cite{weijters2012misresponse}. In that case, even cautious respondents do not understand the antonyms used for item reversal in line with what the researchers intended, which happens especially in cross-cultural research \citep{wong2003cultural}.
Finally, \citet{kam2021people} have highlighted that respondents with a neutral opinion on an issue are likely to choose answers towards a scale's midpoint and tend to agree with both the regular and the reverse-coded items. Given that such respondents have no strong opinion towards either extreme of the scale, this behavior is perfectly reasonable, even if it goes against what the researchers might have expected when constructing the scale \cite{kam2021people}.
As \citet{priester1996gradual} pointed out, the literary figure Hamlet both longs for and, at the same time, fears his death. As a result, he would provide primarily "neutral" or "slightly positive" responses toward suicide on a traditional bipolar attitude scale, which would cause him to agree with both regular and reversed items. 

In summary, using reverse-coded items in questionnaires can result in a two-dimensional scale structure due to the agreement with both regular and reversed items for two reasons.
On the one hand, this contradictory response behavior can happen due to mistakes by the respondents (i.e., careless responding), where the wording of the reverse-coded items is intentionally or unintentionally ignored.
On the other hand, misunderstandings between respondents and researchers (i.e., reversal ambiguity, neutral responses) can also cause response patterns which the researchers did not expect when designing the scale because they expect opposite responses to the reverse and regular items, while respondents agree or disagree with both types. 
As a result of these mistakes and misunderstandings, the regular and reverse-coded items will load on two distinct constructs in factor analysis, not because there are two distinct constructs to be measured but due to methodological issues related to the item wording.

Returning to trust and distrust, it is possible that the two-dimensional structure of the TPA identified by past research \cite{spain2008towards} and in our preliminary findings \cite{perrig2023trust} in the context of AI is not necessarily evidence that trust and distrust are two distinct constructs. Instead, it might indicate the presence of methodological artifacts influencing the measurement of subjective trust in an AI system. Such cases have been reported on in past HCI research and beyond (e.g., psychology \cite{schmitt1985factors}), for example, concerning the System Usability Scale \cite{sauro2011positive} and the Usability Metric for User Experience \citep{lewis2013umux}, both scales which were initially assumed to measure a single uni-dimensional construct (i.e., usability).
Regarding the System Usability Scale, \citet{lewis2017revisiting} recommended still treating the scale as uni-dimensional, measuring just one construct because a distinction due to item tone (i.e., positive or negative) would not make sense based on the underlying theory. A comparable assumption could be made concerning trust measured by the TPA, but in this case, a theoretical distinction between trust and distrust seems more reasonable, as we have illustrated. 

\section{Discussion}

Based on the arguments presented here, we conclude that while a distinction between trust and distrust on a theoretical level appears to be sensible, it is still to be determined if trust and distrust genuinely are two distinct constructs that can be measured independently or if they are the same construct, artificially separated due to methodological issues.

A discussion regarding the role of trust and distrust is relevant because members of the XAI community have emphasized that a key motivation of XAI is to \emph{increase} trust of the user in a trustworthy AI (i.e., warranted trust) but also to \emph{increase} the distrust of the user in a non-trustworthy AI (i.e., warranted distrust) \cite{jacovi2021formalizing}. While we generally agree with this position, the question arises as to how a one-dimensional model of trust can do justice to a simultaneous increase in trust and distrust and the difference between having low trust and distrust. A simplistic understanding of trust may not capture the complexity of peoples' attitudes toward AI. The argument that trust and distrust can coexist simultaneously seems particularly important in today's world, where AI is becoming more and more generalized and can perform multiple tasks (e.g., foundation models). This increased generalizability raises the question of which tasks to trust AI with and which to distrust it with. For example, a large language model (e.g., ChatGPT) might be trusted to write an email but distrusted to generate code or play chess. Consequently, there might be factors that contribute to the increase and decrease of trust, but also factors that contribute to the increase and decrease of distrust \cite{lewicki1998trust}. A two-dimensional model of trust seems more appropriate to account for this circumstance and arguably is more sensitive to such changes, as trust and distrust are not mapped on the same dimension. 

If trust and distrust are distinct, efforts to eliminate distrust do not necessarily establish trust, and in that case, it is necessary to examine whether the two constructs have different antecedents and consequences \cite{chang2013antecedents}. The XAI community might want to give more attention to these differences, and future research could investigate how and if the two constructs can be measured independently in the AI context and what role methodological factors play in this regard. Research outside of an AI context \cite[e.g.,][]{benamati2006trust, chang2013antecedents} attempted to investigate if trust and distrust are distinct constructs in experimental settings. XAI researchers could attempt the same and examine how trust and distrust relate to one another, as well as if the two constructs can predict other subjective and objective measures relevant to the human-AI interaction (e.g., reliance). 

There still seems to be no conclusive answer as to whether trust should be understood uni-dimensionally (with trust and distrust at the two ends of a continuum) or two-dimensionally (with distrust as a distinct and independent construct). However, it is precisely for this reason that this discourse should find its way into the XAI community. In this position paper, we outlined the theoretical arguments in support of a possible distinction between trust and distrust while at the same time showing that this question is not simply answered by psychometrics but by theoretical considerations that feed into experiments. The XAI community should regard these open questions as an opportunity for a more nuanced understanding of the factors influencing the human-AI interaction.

\bibliographystyle{ACM-Reference-Format}
\bibliography{sample}

\end{document}